\documentclass[aps,pre,showpacs,notitlepage,amsmath,amssymb]{revtex4-1}
\usepackage{graphicx}

\begin{document}

\title{$1/f$ noise from the nonlinear transformations of the variables}

\author{Bronislovas Kaulakys}
\email{bronislovas.kaulakys@tfai.vu.lt}
\affiliation{Institute of Theoretical Physics and Astronomy, Vilnius University,\\
A.~Go\v{s}tauto 12, 01108 Vilnius, Lithuania}

\author{Miglius Alaburda}
\email{miglius.alaburda@tfai.vu.lt}
\affiliation{Institute of Theoretical Physics and Astronomy, Vilnius University,\\
A.~Go\v{s}tauto 12, 01108 Vilnius, Lithuania}

\author{Julius Ruseckas}
\email{julius.ruseckas@tfai.vu.lt}
\affiliation{Institute of Theoretical Physics and Astronomy, Vilnius University,\\
A.~Go\v{s}tauto 12, 01108 Vilnius, Lithuania}

\begin{abstract}
The origin of the low-frequency noise with power spectrum $1/f^\beta$ (also
known as $1/f$ fluctuations or flicker noise) remains a challenge. Recently, the
nonlinear stochastic differential equations for modeling $1/f^\beta$ noise have
been proposed and analyzed. Here we use the self-similarity properties of this
model with respect to the nonlinear transformations of the variable of these
equations and show that $1/f^\beta$ noise of the observable may yield from the
power-law transformations of well-known standard processes, like the Brownian
motion, Bessel and similar stochastic processes. Analytical and numerical
investigations of such techniques for modeling processes with $1/f^\beta$
fluctuations is presented.
\end{abstract}

\maketitle

\section{Introduction}

Different theories and models have been proposed for explanation of the
ubiquitous $1/f^\beta$ noise phenomena, observable for about eighty years in
different systems from physics to financial markets (see, e.g.,
Refs.~\cite{Weissman1988,Gisiger2001,Li2012,Balandin2013,Paladino2014,Kaulakys2009,Ruseckas2014,Rodriguez2014}
and references herein). Recently, the stochastic model of $1/f^\beta $ noise,
based on the nonlinear stochastic differential equations
\begin{equation}
dx=\left(\eta -\frac{\lambda}{2}\right) x^{2\eta - 1}dt + x^\eta dW_t\,,
\label{eq:1}
\end{equation}
where $x$ is the signal with $1/f^\beta$ spectrum, $\eta \neq 1$ is the
nonlinearity exponent, $\lambda $ is the exponent of the steady-state
distribution $P_{\mathrm{ss}}\sim x^{-\lambda}$, and $W_t$ is a Wiener process
(Brownian motion), has been proposed and
analyzed.\cite{Kaulakys2009,Ruseckas2014} The relation of the exponent $\beta$
in the spectrum to the parameters of Eq.~(\ref{eq:1}) is given by
\begin{equation}
\beta = 1 + \frac{\lambda - 3}{2(\eta - 1)}\,.
\end{equation}
Eq.~(\ref{eq:1}) may be derived from the point process model
\cite{Kaulakys2009}, from scaling properties of the signal \cite{Ruseckas2014}
or from the agent-based herding model \cite{Ruseckas2011}.

Here we employ the self-similarity property of Eq.~(\ref{eq:1}) with respect the
power-law transformations of the variable. We show that processes with
$1/f^\beta$ spectrum may yield from the nonlinear transformations of the
variable of the widespread processes, e.g., from the Brownian motion, Bessel or
similar familiar processes. 

\section{Transformations}

From the scaling of the power spectral density (PSD) $S(f)\sim f^{-\beta}$,
\begin{equation}
S(af)\sim a^{-\beta}S(f)\,,
\end{equation}
according to the Wiener-Khintchine theorem yields the scaling of the
autocorrelation function
\begin{equation}
C(at)\sim a^{\beta-1}C(t)
\label{eq:3}
\end{equation}
in some time interval $1/f_{\mathrm{max}}\ll t\ll 1/f_{\mathrm{min}}$. Recently
\cite{Ruseckas2014} it was shown that from the scaling (\ref{eq:3}) it follows
the nonlinear stochastic differential equation (SDE)~(\ref{eq:1}). Due to this
scaling the nonlinear SDE
\begin{equation}
dx=\left(\eta_{x}-\frac{\lambda_x}{2}\right)x^{2\eta_{x}-1}dt + x^{\eta_x}dW_t\,,
\end{equation}
generating signal $x$ with PSD
\begin{equation}
S(f)\sim\frac{1}{f^{\beta_x}}\,,\qquad\beta_{x}=
1+\frac{\lambda_{x}-3}{2(\eta_{x}-1)}
\end{equation}
after the nonlinear transformation
\begin{equation}
x=\frac{1}{y^{\delta}}\,,
\label{eq:6}
\end{equation}
with $\delta$ being the transformation exponent, yields SDE for the variable $y$
of the same form,
\begin{equation}
dy=\left(\eta_{y}-\frac{\lambda_y}{2}\right)x^{2\eta_{y}-1}dt + y^{\eta_y}dW_t
\label{eq:7}
\end{equation}
with
\begin{equation}
\eta_y = 1 - \delta(\eta_x -1)\,,\qquad\lambda_y = 1-\delta(\lambda_x -1)\,.
\end{equation}
Eq.~(\ref{eq:7}) generates signal $y$ with the PSD
\begin{equation}
S(f)\sim\frac{1}{f^{\beta_y}}\,,\qquad\beta_{y}=
1+\frac{\lambda_{y}-3}{2(\eta_{y}-1)}=
\beta_x + \frac{1+\delta}{\delta(\eta_{x}-1)}\,.
\label{eq:9}
\end{equation}
Therefore, $1/f^{\beta_x}$ noise of the observable $x$ may yield from the
nonlinear dependence (\ref{eq:6}) of this observable on another variable $y$
resulted from simple or more common equation (\ref{eq:7}), where
\begin{equation}
\beta_x = \beta_y +\frac{1+\delta}{\eta_y-1}\,.
\label{eq:10}
\end{equation}

\section{Examples}

\begin{figure}
\centerline{\includegraphics[width=0.45\textwidth]{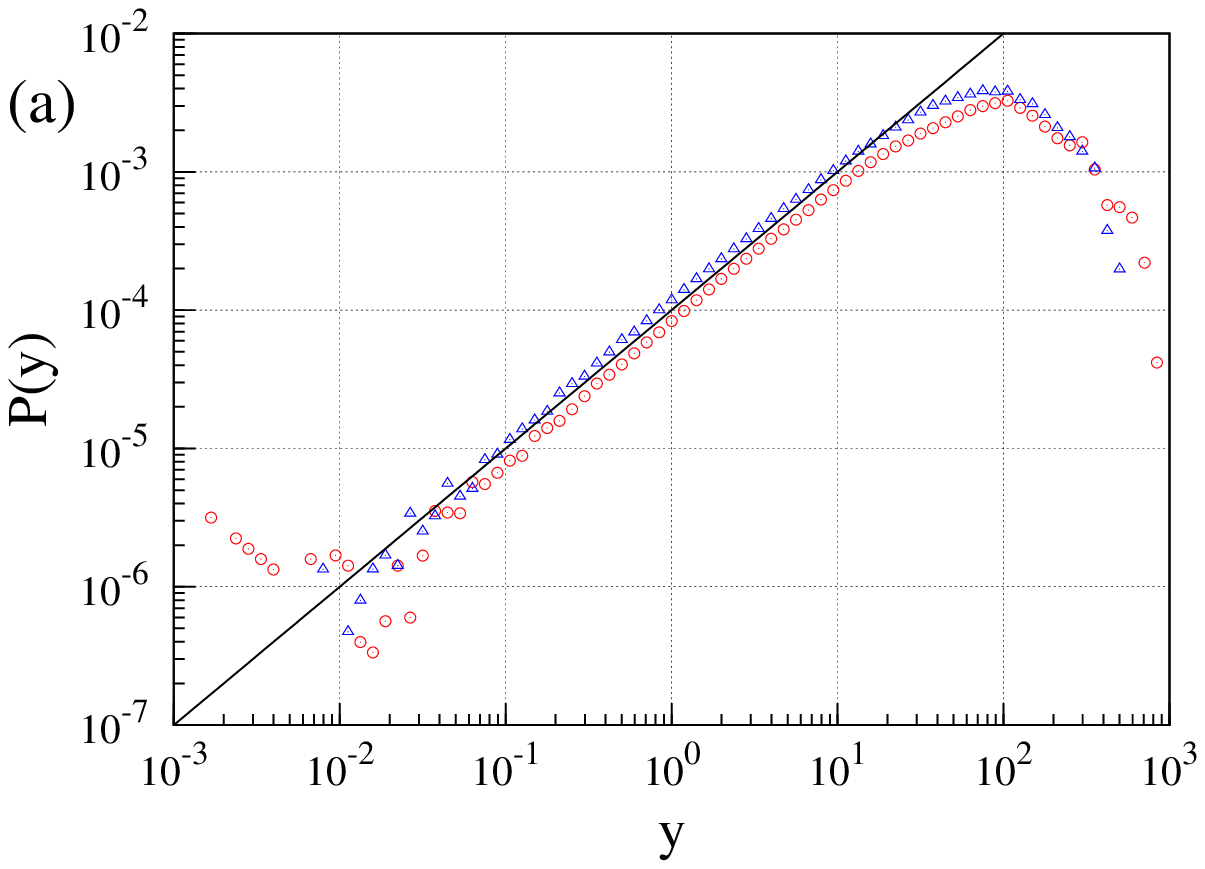}\includegraphics[width=0.45\textwidth]{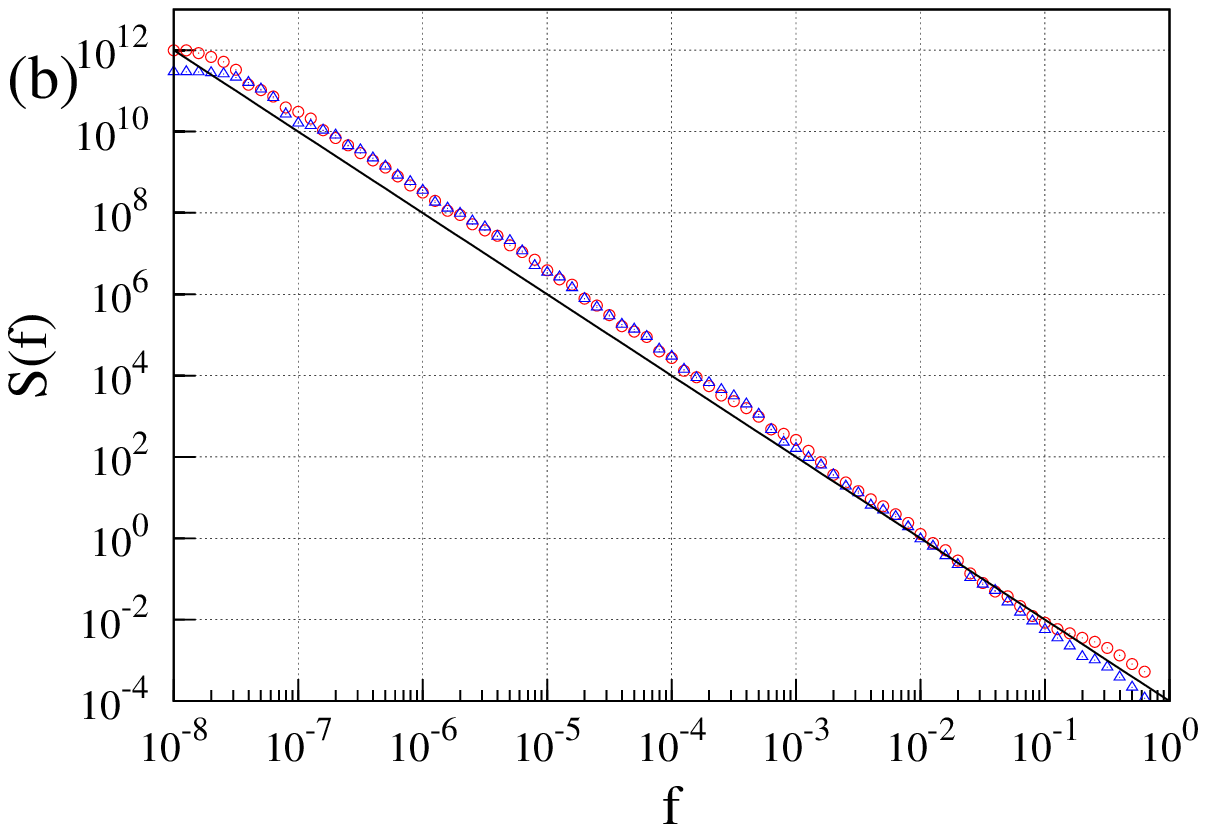}}
\caption{Steady-state distribution, (a), and approximately $1/f^2$ PSD, (b), of
the variable $y$ generated by the simple equation~(\ref{eq:11}), red circles, as
well as of $y=1/x$ generated by Eq.~(\ref{eq:12}), blue triangles, with the
appropriate restrictions\protect\cite{Ruseckas2014} of the diffusion intervals.}
\label{fig:1}
\end{figure}

\begin{figure}
\centerline{\includegraphics[width=0.45\textwidth]{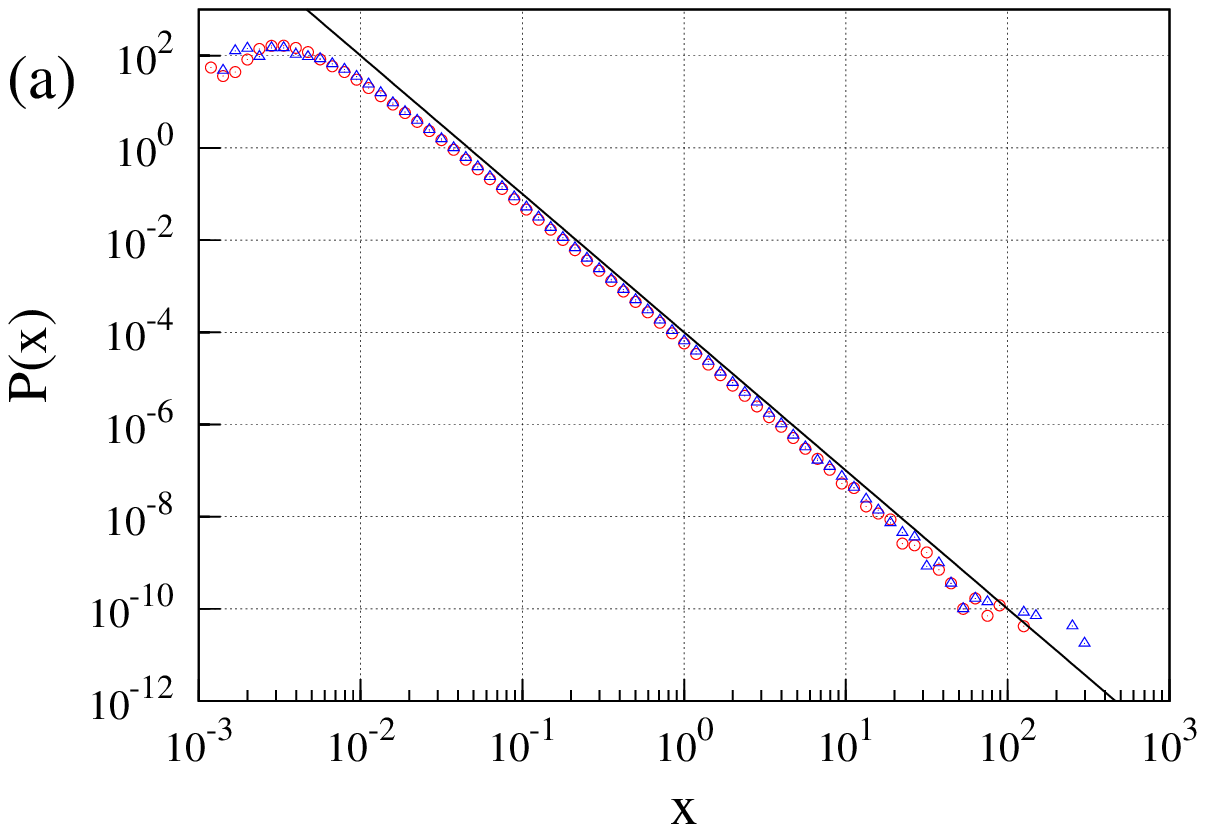}\includegraphics[width=0.45\textwidth]{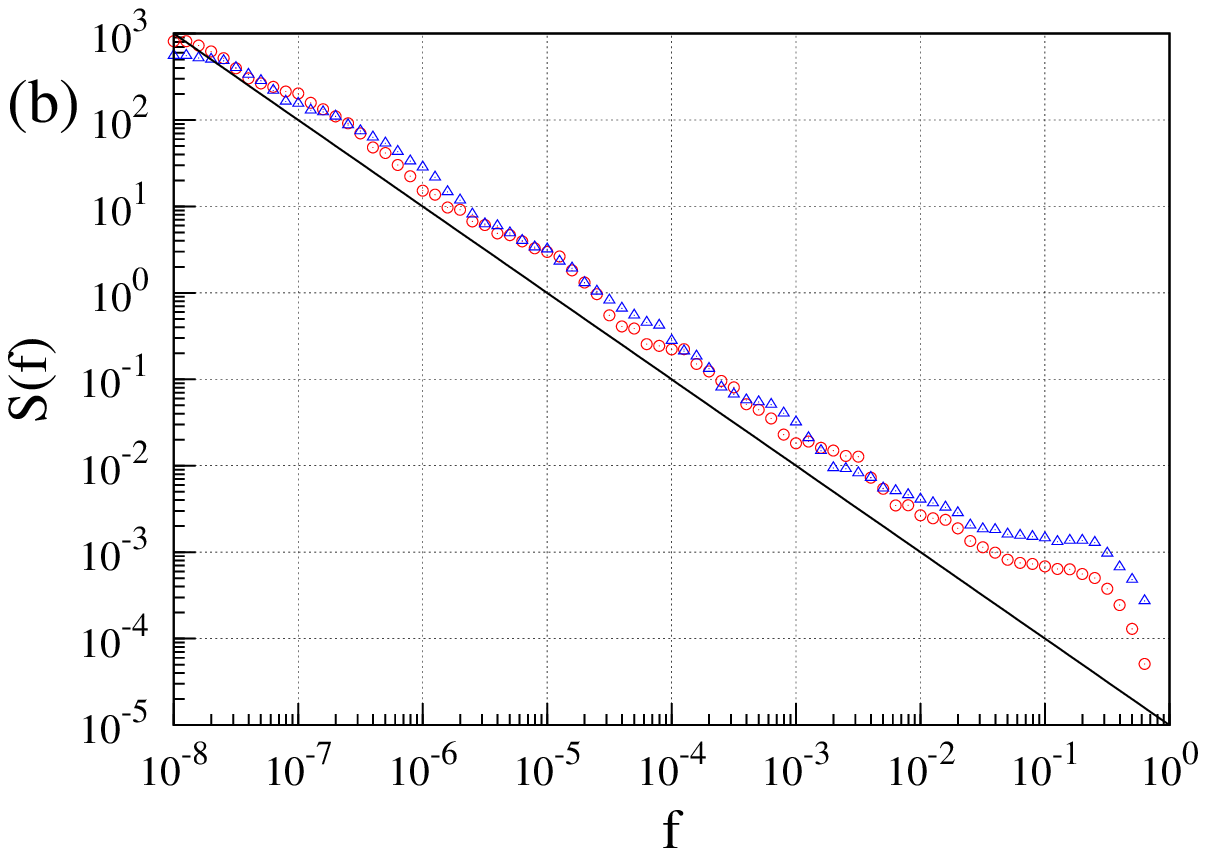}}
\caption{Steady-state distribution, (a), and $1/f$ PSD, (b), of the observable
$x=1/y$, where the variable $y$ is generated by Eq.~(\ref{eq:11}), red circles,
as well as of $x$ generated by Eq.~(\ref{eq:12}), blue triangles.\label{fig:2}}
\end{figure}

\begin{figure}
\centerline{\includegraphics[width=0.33\textwidth]{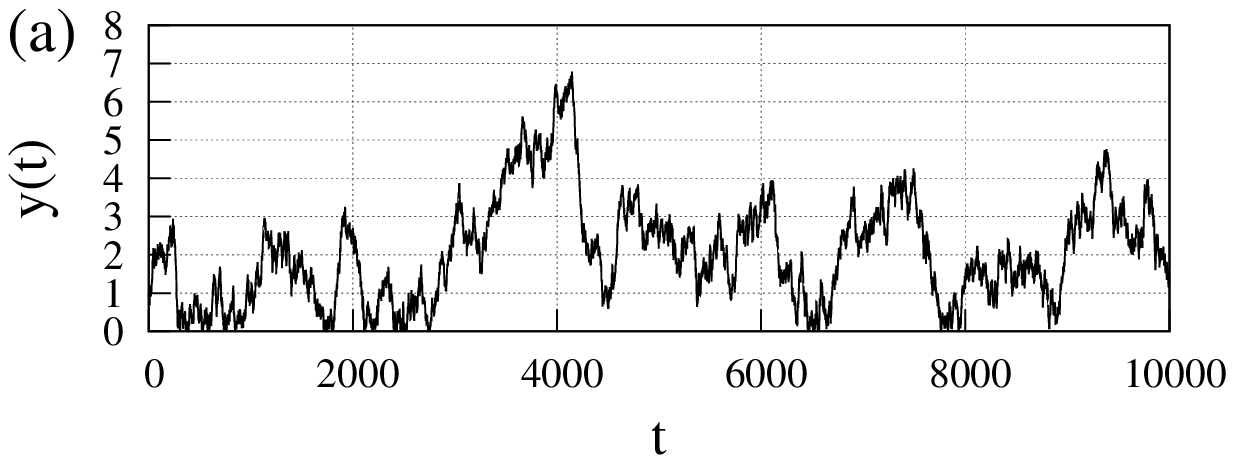}\includegraphics[width=0.33\textwidth]{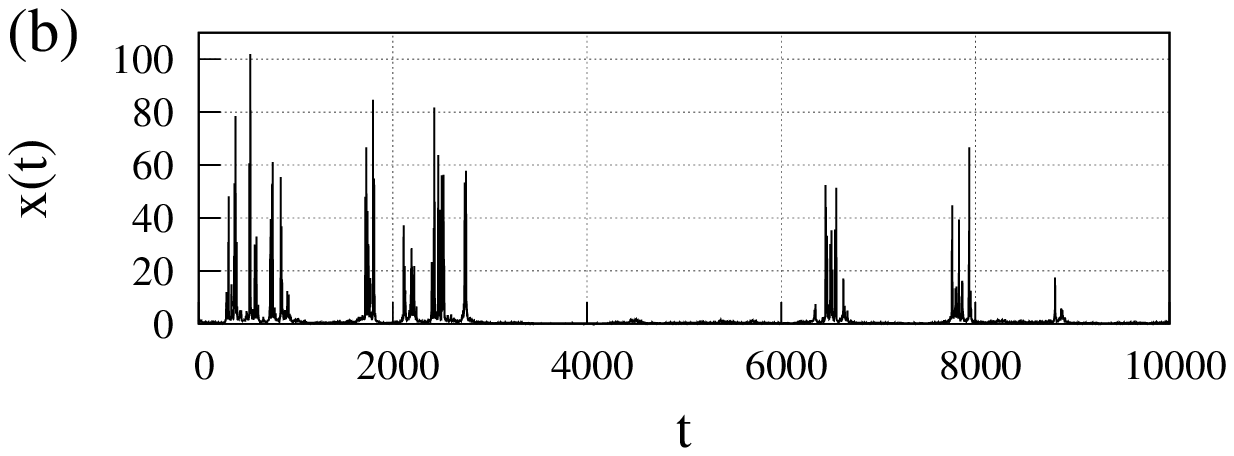}\includegraphics[width=0.33\textwidth]{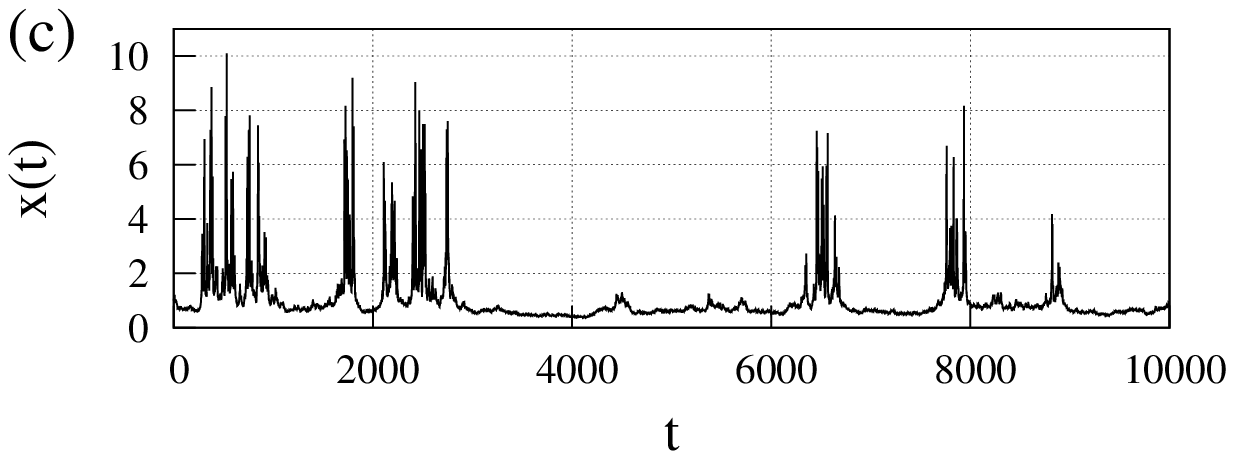}}
\caption{Examples of the signals of 1D Brownian motion of the variable $y$, (a),
of the inverse 1D Brownian motion of the observable $x=1/y$ yielding in
$1/\sqrt{f}$ noise, (b), and of $x=1/\sqrt{y}$ observable resulting in pure
$1/f$ noise, (c).\label{fig:3}}
\end{figure}

\begin{figure}
\centerline{\includegraphics[width=0.45\textwidth]{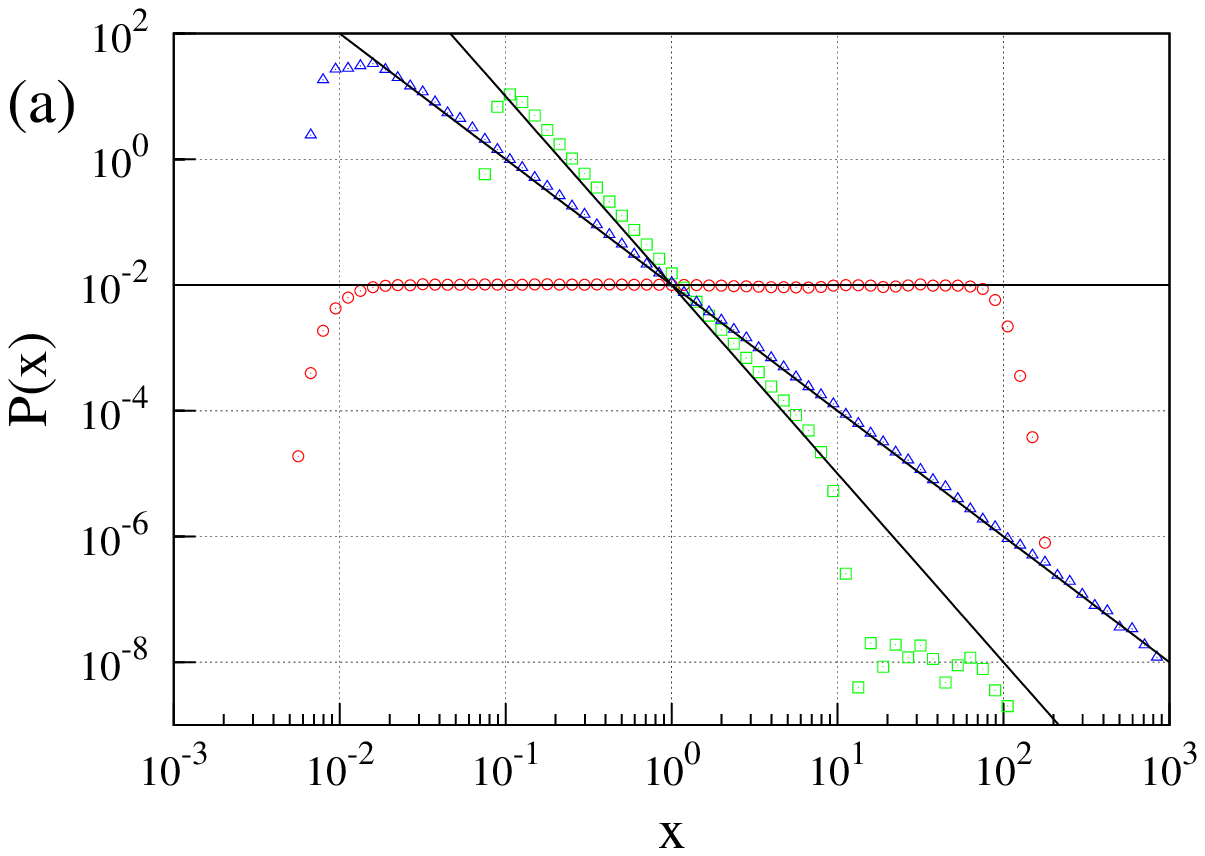}\includegraphics[width=0.45\textwidth]{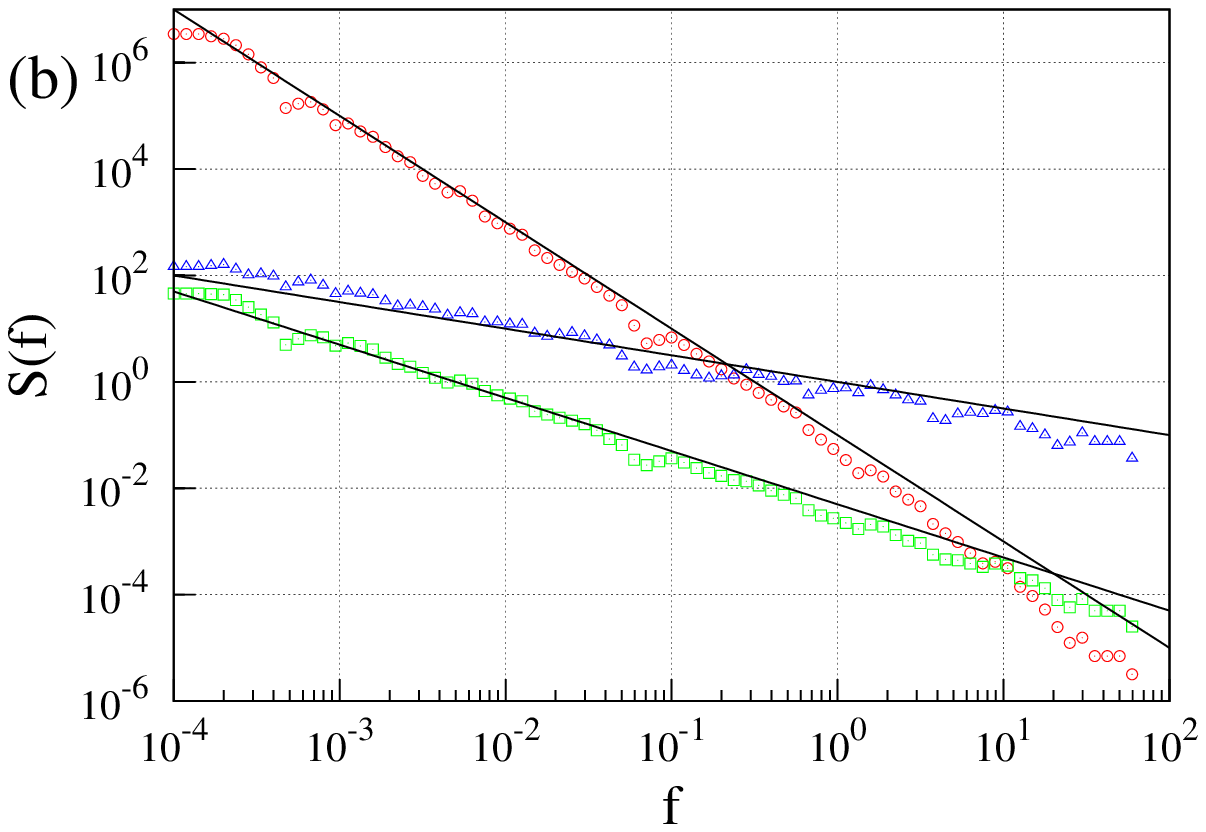}}
\caption{Steady-state distribution, (a), and PSD, (b), of 1D Brownian motion of
the variable $y$, yielding $1/f^2$ noise, red circles, the inverse 1D Brownian
motion of $x=1/y$ yielding in $1/\sqrt{f}$ noise, blue triangles, and of the
observable $x=1/\sqrt{y}$ resulting in pure $1/f$ noise, green
squares.\label{fig:4}}
\end{figure}

\begin{figure}
\centerline{\includegraphics[width=0.45\textwidth]{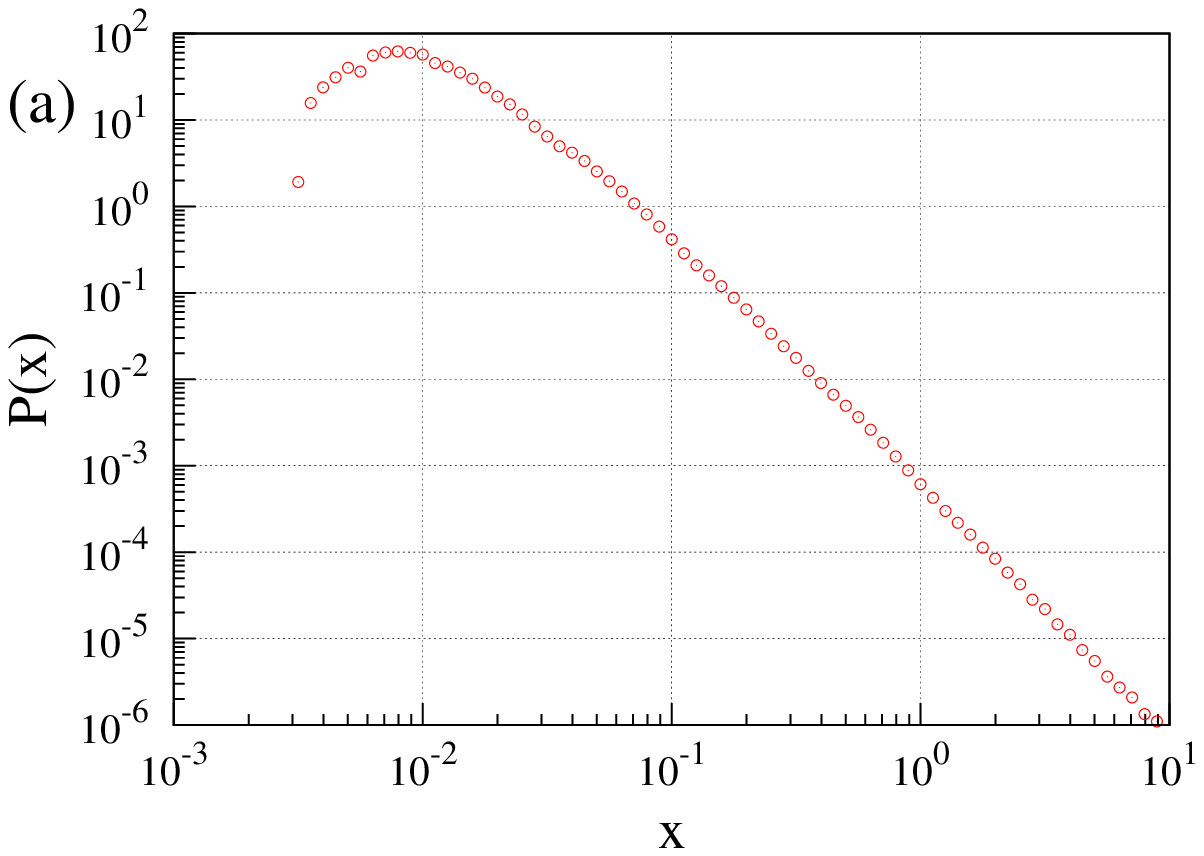}\includegraphics[width=0.45\textwidth]{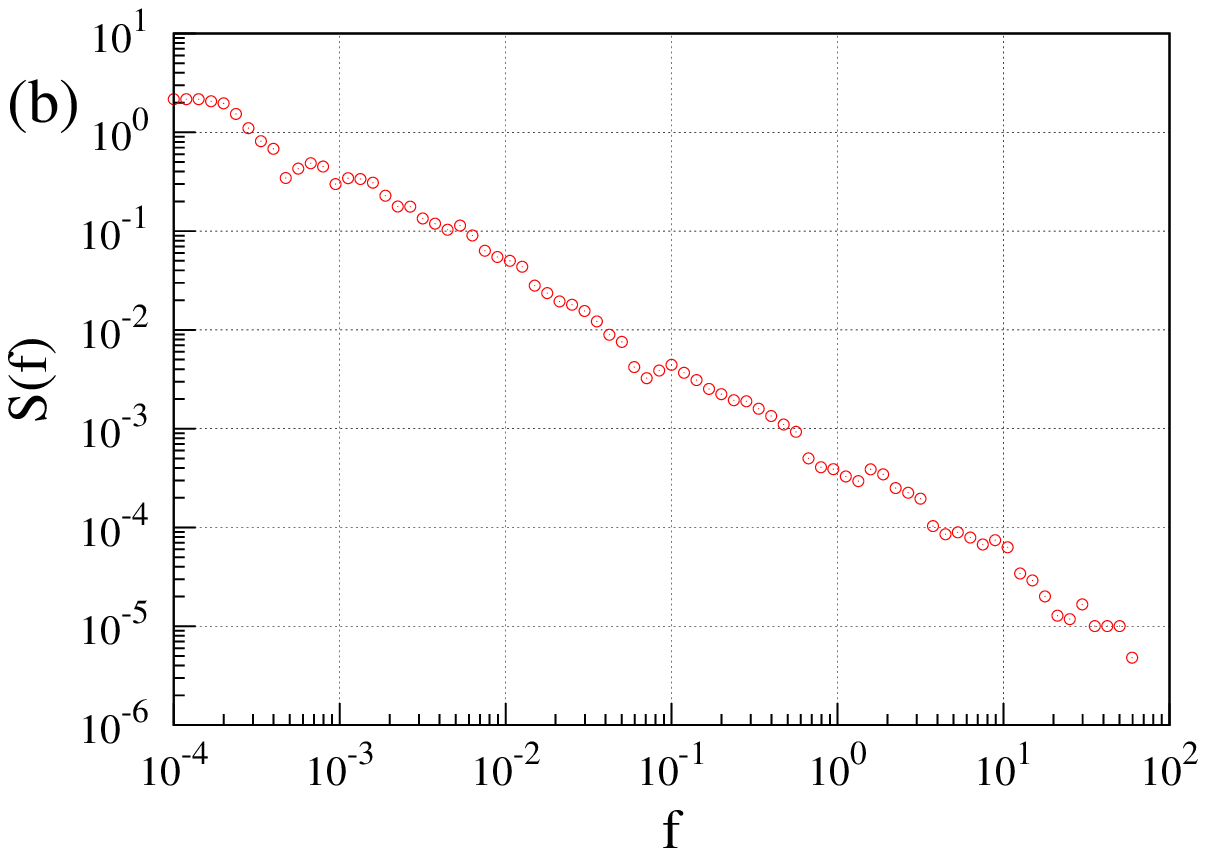}}
\caption{Steady-state distribution, (a), and PSD, (b), of the observable $x$ as
inverse of the distance from the beginning, $x=1/r$, of the Brownian motion in
2D.\label{fig:5}}
\end{figure}

The simplest relation between the variable $y$ and the observable $x$ is the
inverse transformation $x=1/y$. From the interrelations (\ref{eq:9}) and
(\ref{eq:10}) between the parameters $\beta_x$ and $\beta_y$ we obtain that
simple equation without the drift term
\begin{equation}
dy=\frac{1}{\sqrt{y}}dW_t
\label{eq:11}
\end{equation}
results in $1/f$ noise of the observable $x=1/y$, instead of very nonlinear
equation \cite{Kaulakys2004}
\begin{equation}
dx = x^4 dt + x^{5/2} dW_t 
\label{eq:12}
\end{equation}
for the observable $x$. It should be noted, that Eq.~(\ref{eq:11}) coincides
with equation for the interevent time
\begin{equation}
d\tau=\frac{1}{\sqrt{\tau}}dW_t\,,
\end{equation}
obtained form the Brownian motion 
\begin{equation}
d\tau_k = dW_k
\end{equation}
of the interevent time $\tau_k$ in the events space or $k$-space.\cite{Kaulakys2004}

Figures~\ref{fig:1} and \ref{fig:2} demonstrate the appearance of $1/f$ noise of
the observable $x$, as a result of the inverse transformation $x=1/y$ of the
variable $y$ generated by the simple equation without the drift
term~(\ref{eq:11}) with approximate $1/f^2$ PSD of the variable $y$.

Note, that special cases of Eq.~(\ref{eq:7}) are: (i) an equation with the
additive noise, $\eta_y =0$, and nonlinear drift, i.e., the Bessel process of
order $N=1-\lambda_y$, (ii) the order $N=2(1-\lambda_y)$ squared Bessel process
when $\eta_y=1/2$ and $\lambda_y=1-N/2$, (iii) the special exponential
restrictions of the variable $y$ in Eq.~(\ref{eq:7}) yield Constant Elasticity
of Variance (CEV) process or (iv) Cox-Ingersoll-Ross (CIR) process when
$\eta=1/2$ \cite{Ruseckas2011,Jeanblanc2009}.

Here we will demonstrate the possibility to obtain $1/f^{\beta}$ noise from the
Bessel process and from the Brownian motion. Transformation (\ref{eq:6}) with
$\delta=1/(\eta_x -1)$, i.e., $y=x^{1-\eta_x}$, yields the special form of
Eq.~(\ref{eq:7}), i.e., the Bessel equation
\begin{equation}
dy=\frac{N-1}{2}\frac{dt}{y}+dW_t
\end{equation}
corresponding to $N$-dimensional Wiener process, or Bessel process with index
$\nu=N/2-1$. Here
\begin{equation}
N=\frac{\lambda_x -1}{\eta_x -1} = 1-\lambda_y\,.
\end{equation}
On the other hand
\begin{equation}
\lambda_x = 1 +\frac{N}{\delta}\,,\qquad\beta_{x}=1+\frac{N}{2}-\delta\,.
\end{equation}
Thus, the Bessel process or $N$-dimensional Brownian motion can cause
$1/f^{\beta}$ fluctuations of the observable $x$ as a function (\ref{eq:6}) of
this process. So, the pure $1/f$ noise yields when $N=2\delta$, e.g., from 1D
Brownian motion of $y$ for the observable $x=1/\sqrt{y}$ and 2D Brownian motion
of $y$ for the observable $x=1/y$.

Figure~\ref{fig:3} demonstrates examples of the signals of 1D Brownian motion of
the variable $y$, the inverse 1D Brownian motion of $x=1/y$ yielding in
$1/\sqrt{f}$ noise and observable $x=1/\sqrt{y}$ resulting in pure $1/f$ noise.
In figure~\ref{fig:4} the distribution density and PSD of the corresponding
observables are presented. 

In figure~\ref{fig:5} the pure $1/f$ noise of the observable $x$ as inverse of
the distance from the beginning, $x=1/r$, of the Brownian motion in 2D, 
\begin{equation}
r=\sqrt{B_1^2\left( t\right) +B_2^2\left( t\right) }\,, 
\end{equation}
i.e., of the Bessel process with the index $\nu=0$ is shown. Here $B_1\left(
t\right)$ and $B_2\left( t\right)$ are independent Brownian motions. We see the
pure $1/f$ noise. 

\section{Conclusions}

Thus, the proper function of the widespread process, e.g., the inverse
transformation of the order 2 Bessel process and the inverse square root of the
Brownian motion yield the pure 1/f noise, observable, e.g., in condense matter.
Possible relevancies of these transformations to $1/f$ noise in condensed
matter: (i) $1/f$ fluctuations of the voltage, $U=I/G$, at constant current $I$
as a result of Brownian fluctuations of the conductivity $G$, (ii) $1/f$
fluctuations of resistivity, $\rho =1/\sigma$, as Brownian fluctuations of the
conductivity $\sigma$.

\end{document}